
\newskip\oneline \oneline=1em plus.3em minus.3em
\newskip\halfline \halfline=.5em plus .15em minus.15em
\newbox\sect
\newcount\eq
\newbox\lett

\def\simlt{\mathrel{\lower2.5pt\vbox{\lineskip=0pt\baselineskip=0pt
           \hbox{$<$}\hbox{$\sim$}}}}
\def\simgt{\mathrel{\lower2.5pt\vbox{\lineskip=0pt\baselineskip=0pt
           \hbox{$>$}\hbox{$\sim$}}}}

\newdimen\short
\def\adv{\global\advance\eq by1}
\def\set#1#2{\setbox#1=\hbox{#2}}
\def\nextlet#1{\global\advance\eq by-1\setbox
                \lett=\hbox{\rlap#1\phantom{a}}}

\newcount\eqncount
\eqncount=0
\def\equn{\global\advance\eqncount by1\eqno{(\the\eqncount)} }
\def\put#1{\global\edef#1{(\the\eqncount)}           }

\magnification=1200
\hsize=6.0 truein
\vsize=8.5 truein
\baselineskip 14pt

\nopagenumbers
\rightline{hep-ph/9403348}
\rightline{CPTH-A292.0294}
\rightline{February 1994}
\vskip 1.0truecm

\centerline{\bf  $\pi \pi$ SCATTERING AND PION FORM FACTORS}
\vskip 1.0truecm
\centerline{{\bf L. Beldjoudi} and {\bf Tran N. Truong}}
\vskip .5truecm
\centerline{{\it Centre de Physique Th{\'e}orique}
\footnote{$^*$}{\it Laboratoire Propre du CNRS UPR A.0014}}
\centerline{\it Ecole Polytechnique, 91128 Palaiseau, France}
\vskip 2.5truecm
\centerline{\bf ABSTRACT}
\vskip .5truecm

The S and P wave $\pi \pi$ phase shifts are recalculated in terms of two
phenomenological
parameters using the one loop CPTh
and the elastic unitarity condition.
Using these phase shifts, the vector and scalar form factors are calculated
and  shown to be in a  good agreement with experimental data. It is found
that the simpler and more phenomenological approach, where the  left hand
cut contributions to the partial wave amplitude are neglected, yields
approximately
the same result.

\vskip 3.5truecm

\vfill\eject

\footline={\hss\tenrm\folio\hss}\pageno=1

\vskip 1.5 cm

Chiral Perturbation Theory (CPTh) is a low energy expansion in the
Nambu-Goldstone boson momenta which takes into account in a systematic manner
the chiral symmetry breaking effect [1]. The matrix elements obtained from CPTh
can be considered as approximate low energy theorems. They have been
successful in correlating a number of low energy phenomena involving soft
pions.

Unfortunately, these low energy theorems are only valid near  the threshold
region and cannot take into account of the resonance phenomenon or of a very
strong interaction. The basic reason why it cannot handle these situations
is because it is a perturbative calculation and  can only satisfy
perturbatively the
unitarity relation. In contrast, the low energy effective range theory of
the nucleon
nucleon scattering, which satisfies the elastic unitarity exactly can
handle both the
bound state and resonance problems. We therefore want also to impose the
exact elastic unitarity
relation for the $\pi\pi$ partial wave amplitudes. It is then not difficult
to make a
resummation of the one loop CPTh into a geometric like series so that the
elastic unitarity condition is exactly satisfied. This geometric like series
can be constructed from the one loop calculation and is known as the Pad\'{e}
[1,1] approximant method. We use here the Pad\'{e} method as a most
convenient way to
satisfy exactly the elastic unitarity relation, but not in the mathematical
sense
of summing a diverging series. The use of the Pad\'{e} method in order to
satisfy the elastic unitarity relation is
certainly not unique. It has been shown,  for the special case of the P
wave $\pi \pi$ scattering, all methods lead to essentially the same physical
consequences [2], namely, given the correct scattering length and the
effective range, the
$\rho$ resonance  exists once the elastic unitarity condition is
implemented.

The purpose of this note is to recalculate the well known one loop CPTh for
 $\pi \pi$
scattering. The partial wave
amplitudes  are then resummed by the Pad\'{e} approximant method in order
to satisfy
the elastic unitarity condition. We then
calculate the S and P waves phase shifts to be used in the Omn\`{e}s
representation [3] in order to calculate the corresponding pion form factors.

 We want to point out that this more exact calculation of the form factors
yields
essentially the same result as the simplified approximation of the N/D
method where the
discontinuity of the left hand cut contribution is neglected.

Because we do not use the standard CPTh to calculate the form factors [4], our
calculation of the form factors, using Omn\`{e}s representation, involves the
same  parameters as those used in the scattering
amplitudes and therefore no new parameters are needed. In particular we can
directly calculate the pion  r.m.s radii in
terms of the calculated phase shifts and compared them with the
experimental data.

\vskip 1.0 cm
{\bf I) Calculation of pion pion scattering phase shifts}
\vskip 0.5 cm

 The most general $\pi^{a}(p_{1})\pi^{b}(p_{2})
\to\pi^{c}(p_{3})\pi^{d}(p_{4})$ scattering amplitude  is given by:

$$A(s,t,u)\delta^{ab}\delta^{cd}+A(t,s,u)\delta^{ac}\delta
^{bd }+A(u,t,s)\delta^{ad}\delta^{bc} \eqno (1)$$

In the $\pi \pi$ scattering we have three isospin amplitude $I=0,1,2 $ which
are
simple functions of A(s,t,u):

$$\eqalign{ & T_{0}(s,t,u)  = 3 A(s,t,u)+ A(t,s,u)+A(u,t,s)\cr
&
T_1(s,t,u) = A(t,s,u)-A(u,t,s)\cr
&
T_2(s,t,u) = A(t,s,u)+A(u,t,s) }\eqno (2) $$

Where s,t,u are Mandelstam variables with $s+t+u=4m_{\pi}^2$. Without loss of
generality, they can be decomposed into the partial waves:

$T_I(s,t)=32\pi \sum_{l}(2l+1)P_l(cos\theta)t_{l}^I(s) $

Below the inelastic thresholds such as $K\bar{K}$,  $\pi \omega$,  the
unitarity of the S matrix
implies Im$t_{l}^I(s) =\sigma (s)\vert t_{l}^I(s) \vert^2 $ for $s \ge
4m_{\pi}^2$
where $\sigma (s)=\sqrt{ 1-{4m_{\pi}^2 \over s} } $ is the phase space factor.
Because of the unitarity relation, the partial wave amplitude can be written
as $t_{l}^I(s)= {e^{+i\delta_l^I} \sin(\delta_l^I) \over \sigma (s) }
$

At  one loop chiral perturbation theory [5,6],  we have:

$$A(s,t,u)= {s-m_{\pi}^2 \over f_{\pi}^2 }+B(s,t,u)+C(s,t,u)+0(p^6) \eqno
(3-a) $$

where

$$\eqalign { B(s,t,u) &={1\over 6f_{\pi}^4} [ 3(s^2-m_{\pi}^4)\bar{J}
(s)+( t(t-u)-2m_{\pi}^{2}t+4m_{\pi}^{2}u-2m_{\pi}^4)\bar{J} (t)\cr
&
+( u(u-t
)-2m_{\pi}^{2}u+4m_{\pi}^{2}t-2m_{\pi}^4)\bar{J} (u)] } \eqno (3-b)$$

and

$$\eqalign {C(s,t,u) &={1\over 96 \pi^2 f_{\pi}^4}[ 2(\bar
{l_1}-4/3)(s-2m_{\pi}^{2})^2+(
\bar {l_2}-5/6)(s^2+(t-u)^2)\cr
&
+12m_{\pi}^{2}s(\bar {l_4}-1)-3(\bar {l_3}+4\bar
{l_4}-5 )]} \eqno (3-c)$$

These results are obtained  from the $0(p^4)$ calculation using $SU(2)_L
\times SU(2)_R$ chiral
 lagrangian:

$$\eqalign{L &={F^2\over 4}tr\partial_{\mu}U\partial^{\mu}U^{\dagger}+
{F^2\over 4}{m^2}tr(U+U^{\dagger})
+{l_1 \over 4}tr(\partial_{\mu}U\partial^{\mu}U^{\dagger})^2\cr
&
+{l_2 \over
4}tr(\partial_{\mu}U\partial^{\nu}U^{\dagger})tr(\partial_{\mu}U\partial^{\nu
}U^{\dagger})+{l_3+l_4 \over 16}m^4 (tr(U+U^{\dagger}))^2\cr
&
+{l_4 \over 8}m^2 tr(\partial_{
\mu}U\partial^{\mu}U^{\dagger})tr(U+U^{\dagger})}  \eqno (4)   $$  where
$U=\exp  {i\pi ^{a} \tau ^{
a} \over F}$ is the exponential representation of the pion field, m and F
are the bare mass and
decay constant of the pion.

In Eq.(3) $f_{\pi}=93.3$ Mev is the physical pion decay constant, up to a
scale factor
 $\bar {l_i}$ are
function of $ l_{i}^r(\mu)$ plus a quantity which make them scale independent
as indicated in ref. [6].   The constants $\bar {l_i}$ are scale
independent parameters. $\bar{J} (t)
$ is the once substracted scalar two point function.
\vskip 0.5cm

$$\bar{J} (t)={1\over 16{\pi}^2}(2+\cases {\sigma( \log ({1-\sigma \over
1+\sigma}) +i\pi )
& for $t\ge 4m_{\pi}^2$\cr
\sigma \log ({-1+\sigma \over 1+\sigma})
)& for $t \le 0$\cr
-2 \vert \sigma \vert \arctan {
1\over\vert \sigma \vert} ) & otherwise}) \eqno (5)$$.

\vskip 0.5cm

Below $K\bar{K}$ production the one loop chiral perturbation theory
satisfies the
perturbative unitarity:  Im $t^{(1)}=\sigma (s) {t^{(0)}}^2 $ where $
t=t^{(0)}+
t^{(1)} $ and the superscripts stand for the tree graph and one loop
calculation; the isospin index are omitted for convenience. It is
staightforward
to show that the reconstructed amplitude

$$ t(s)={t^{(0)}\over 1-{t^{(1)}\over t^{(0)} } } \eqno (6)$$ satisfies
exactly the elastic
 unitarity. Our determination of $\bar{l_{i}}$ parameters is different than
ref.[6],
where one can deduce
$\bar {l_{1}}$ and $\bar {l_{2}}$ from  I=0,I=2  D wave scattering
length. It was found that $ \bar {l_{1}}=-0.6$, and $\bar {l_{2}}=6.3$. In our
calculation the quantity $\bar {l_{2}}-\bar {l_{1}}$ is sensitive to the $\rho$
resonance and it is directly fixed by the $\rho$ mass. The $\rho$ mass is
defined as the energy
 where the I=1,  {l}=1 phase shift passing through
$90$ degrees. To establish completely  the unknown parameters, we use the
experimental
 S wave I=0 phase shift at 500 Mev [7]. $\bar {l_{3}}$ and $\bar {l_{4}}$
measure the chiral symmetry breaking
effect; their contributions to the scattering amplitude are proportional to
the pion mass
squared. In this calculation they are given by the reference [6]  $\bar
{l_{3}}=2.9$ $\bar {l_{4}}=4.3$, which are determined by the
SU(3) mass formula and the ratio $f_{K}/f_{\pi}$.  Our best values are $ \bar
{l_{1}}=-0.45$, and $\bar {l_{2}}=5.51$, which are slightly different from
those determined
by[8]. The $\rho$ mass is
taken to be its experimental value $m_{\rho}=770 Mev$.

Once we fix the $\bar {l_{i}}$ parameters $\pi \pi$ scattering amplitude is
known. The predictions
to the scattering length are:
$a_{0}^{0}=0.22$, and $a_{1}^{1}=0.038$. These values are consistent with
those given by experiments. In Fig[1-2],  S and P wave phase
shifts are compared with experimental data. It can be seen that the
agreement between the theoretical
predictions and experimental data are exellent. The deviation of the
theoretical prediction and
experimental data for I=0 S wave above $700$ Mev is due to the opening of
the inelastic channel
$ a^{0} (980)$ which is not included in our calculation.

The diagonal [1,1] Pad\'{e} method for S waves is not without a problem:
because $t^{(0)}$ has a zero at
$s={m_{\pi}^2\over 2}$, from Eq (6), it is clear that the denominator of
$t(s)$ has a zero near
${m_{\pi}^2\over 2}$ or t(s) has a spurious pole near ${m_{\pi}^2\over 2}$.
It is simple to show
that it has a very small residue. We must, in principle, substract this
pole from t(s) in order to
have a correct analytical property. Such a substraction has a little on the
calculated
amplitude and would result in a tiny violation of the
unitarity relation in the physical region. We shall ignore, in the
following this substraction procedure.

\vskip 1.0 cm
 {\bf II) Calculation of the pion form factors}
\vskip 0.5 cm
In the remainder of this paper, we calculate the vector and the scalar form
factors.
Using the CVC hypothesis and Lorentz invariance, one can write
straightforwardly
the following matrix elements:
$$\eqalign { &\langle\pi^{a}(p_1) \pi^{b}(p_2)\vert V_{\mu}^{c} (0)\vert0
\rangle =i\epsilon ^{abc}f(s)(p_2-p_1)_\mu  \cr
&\langle\pi^{a} \pi^{b}\vert \hat {m}(\bar {u}u+\bar {d}d)\vert0
\rangle =\delta^{ab}\Gamma (s)  } \eqno (7)$$

At zero momentum transfer  we have the following normalization: $f(0)=1$ and
$\Gamma (0)=m_{\pi}^{2}$.
  Assuming the elastic
unitarity condition, we deduce the following relations:

$$ Imf(s)=f(s)\exp {-i\delta_1^{1} }\sin {\delta_1^{1}} \eqno (8-a) $$
$$ Im\Gamma(s)=\Gamma(s)\exp {-i\delta_0^{0} }\sin {\delta_0^{0}} \eqno (8-b)
$$
Hence $f(s)$ must have the phase of the P wave phase shift, and $\Gamma(s)$ the
S wave phase shift.

The general solutions to this equation are well known, they are of the
Muskhelishvili Omnes type:
$$\eqalign {& f(s)=P_{f}(s)\Omega_1 (s)\cr
&\Gamma(s)=\Gamma(0)P_{\Gamma}(s)\Omega_0 (s) } \eqno (9)$$
where

$$\eqalign{ &\Omega_1 (s)=\exp ({{s\over \pi}\int\limits_{4m_\pi^2} ^{+\infty}
{
\delta_1^{1}dz\over z(z-s-i\epsilon}}) \cr
&
\Omega_0 (s)=\exp ({{s\over \pi}\int\limits_{4m_\pi^2} ^{+\infty} {
\delta_0^{0}dz\over z(z-s-i\epsilon}}) } \eqno (10) $$

$P_{f}$ and $P_{\Gamma}$ are polynomials which  determine the high energy
behavior of the form factors. They could also represent the low energy
contribution of the higher
mass intermediate states to the form
factors. In the following we assume the dominance of
the elastic unitarity relation and hence we set  $ P_{f}(s)=P_{\Gamma}(s)=1$.

Using the S and P wave phase shifts as calculated above, the scalar and
vector form factors
are calculated numerically, using Eqs(9). Because there are no experimental
informations on the
scalar form factor, we only compare the vector form factor with the
experimental data. It is
seen that the agreement between theory and experimental data is
satisfactory although the peak
values of the experimental form factor squared at the $\rho$ mass is about
40 as compared with
the theoretical calculation value 32 or an error of the order of 20\%. This
discrepancy is
probably due to the inelastic effect of the $\omega \pi$ channel as was
previously pointed
out [9].

We can also calculate the vector and scalar pion rms radii using the
following formula:
 $$\langle r^2_{V}\rangle={6\over \pi} \int\limits_{4m_\pi^2} ^{+\infty} {
\delta_1^{1}dz\over z^2} $$

$$\langle r^2_{S}\rangle={6\over \pi} \int\limits_{4m_\pi^2} ^{+\infty} {
\delta_0^{0}dz\over z^2 }$$
Numerical integration gives $\langle r^2_{V}\rangle=
0.40 fm^2$, and  $\langle r^2_{S}\rangle = 0.47 fm^2$ compared to the
experimental value $\langle r^2_{V}\rangle = 0.439\pm 0.03 fm^2$,
and the $\langle r^2_{S}\rangle=0.5 fm^2 $ which is obtained from the
experimental $\pi K$
scalar radius and from SU(3) symmetry. The agreement between experimental
data and theoretical
 calculations is  satisfactory.

{\bf III) Phenomenological approximations}

The following approximation schemes have been used in the litterature
[10-11-12]. Decomposing the
partial wave amplitudes as $t_{l}^{I}=N_{l}^{I}/D_{l}^{I}$ where for
convenience we normalize
$D_{l}^{I}(0)=1$.  In this case $D_{l}^{I}$ is identical to the function
${\Omega}_{l}^{I}$ defined
in Eq(10).
Approximating $N_{l}^{I}$ by the tree amplitude $N_{l}^{I}(s)=t^{0}(s)$, we
have:
$$D_{l}^{I}(s)=1+b_{l}^{I} s-{s^2\over \pi}
\int \limits_{4m_{\pi}^2} ^{+\infty}{\sigma (z)N_{l}^{I}(z) dz\over
z^2(z-s-i\epsilon)} \eqno (11) $$
This N/D construction satisfies the elastic unitarity condition exactly but
it neglects the
contribution from the one loop graph in t and u channel. $b_{l}^{I}$'s are
phenomenological parameters to be determined from the experimental data.

 How good are these
approximations and what are their relation with the unitarized Chiral
Perturbation Theory?

By explicit numerical evaluation of the expression correspending to the
left hand cut contribution
in the denominator of Eq[6] one can show that it can be approximated with a
good accuracy by
a polynomial for $4m_{\pi}^2\le s \le 1Gev^2 $. The validity  of this
approximation explains why
the left hand cut contribution can be absorbed into the adjustable parameter
$b_I^l$ in the
$D_I^l$ function defined by Eq(11).

The P wave result in this approximation is well known as first given by
Brown and Goble [10]. Fitting
the unknown $b_{1}^{1}$ with the $\rho$ mass, defined as the energy where
the P wave phase shift
passing through 90 degrees, we get the usual KSRF relation giving the
$\rho$ width $\Gamma _{\rho}=
142$ Mev which is smaller than the experimental value $\Gamma _{\rho}=153$
Mev. Our exact calculation,
taking into account the left hand cut yields $\Gamma _{\rho}=157$ Mev which
is in excellent agreement
with the data. The modulus squared of the pion form factor calculated in
this approximation is
shown in Fig[3]. It is seen that it is in a good agreement with that given
by the exact
calculation. Because the difference between these two calculations is
small, we conclude that the
neglect of the left hand cut discontinuity coming from t and u channel one
loop graph, and treating $b_I^l$
as an adjustable parameter is a good approximation.

 Similarly we make the same
approximation for the I=0 scalar form factor. In Fig [1-4]
we compare the phase and modulus of the approximate calculation, where the
t and u one loop
contibution is neglected with those where their effects are taken into
account. It is seen that
there  is little difference between the approximate and the exact results.

The expression for the I=0 S wave D function $D_{0}^{0}$ without the
contribution of the left hand
cut to the scattering amplitude was previously used in connection with the
$K \to 2\pi$, $K\to 3\pi$, $K_{S}\to 2\gamma$, $\gamma
\gamma \to 2\pi $ and $K_{L}\to {\pi}^{0}\gamma\gamma $ problems. Various
results given in ref[12] remain of course valid when the more exact
calculation
of $1/D_{0}^{0}$ function is used.

We show in this article the interrelation between the scattering amplitudes
and form factors.
Instead of using CPTh in the low energy region near to $2\pi$ threshold, we
extend its validity
to a much wider energy region by imposing the unitarity relation to take
into account of
the resonance effect (in P wave). It has recently shown that CPTh for
the scalar form factor, using some prescriptions, can be extended to
400-500 Mev, and can
 describe $K_{S}\to 2\pi$ amplitude with the correct phase[13]. It remains
to be seen whether the same
prescriptions can be used without ambiguity for the $\gamma\gamma \to
2\pi$, $K_{S}\to
2\gamma$, and $K_{L}\to \pi\gamma
\gamma$ problems.

\vskip 1.0 cm
\centerline {\bf {REFERENCES}}
\vskip 0.5 cm

\parskip=-5 pt
\item{[{1}]} S.Weinberg, {\it Physica A} {\bf 96 }(1979) 327.\hfill\break

\item{[{2}]} T.N Truong, {\it Phy. Rev. Lett.}{\bf 67},(1991)2260.\hfill\break
\item{[{3}]} N.I.Muskhelishvili, {\it Singular integral equations
(Noordhoff, Groningen,1953)};

R. Omnes, {\it Nuovo Cimento} {\bf 8 }(1958)316.\hfill\break

\item{[{4}]} J. Bijnens and F. Cornet, {\it Nucl.Phy.}B{\bf 296}
(1988)557.\hfill\break

\item{[{5}]} K.S. Jhung and R.S. Willey, {\it Phy. Rev.}D{\bf 9} (1974)
3132.\hfill\break

\item{[{6}]} J.Gasser and H. Leutwyler, {\it Ann.Phy. (NY)}{\bf 158 }(1984)
142;

Nucl.Phy.B{\bf 250} (1985)465.\hfill\break

\item{[{7}]} P. Estabrooks and A.D. Martin, {\it Nucl.Phy.} B{\bf 79}
(1974)301.\hfill\break

\item{[{8}]} A. Dobado and J.R Pelaez Z.Phy. C{\bf 57} (1993)501.\hfill\break

\item{[{9}]} B. Costa de Beauregard, T.N.Pham, B.Pire and T.N. Truong, {\it
Phy. Lett.}B{
\bf 67} (1977)213.\hfill\break
\item{[{10}]} L.S. Brown and R.L.Goble, {\it Phy. Rev. Lett.} {\bf 20},
(1968)346.\hfill\break

\item{[{11}]} T. N. Truong, {\it  Phy. Rev. Lett.},{\bf 61}(1988)2526.
\hfill\break
\item{[{12}]} T.N. Truong, {\it Phy. Lett.} B{\bf 314} (1993)217.\hfill\break
\item{[{13}]} J. Gasser and Ulf G. Mei${\beta}$ner Bern University preprint
BUTP-90/20.
\hfill\break
\item{[{14}]} L.M. Barkov et al, {\it Nucl.Phy.}B{\bf 256}(1985)365

 \vskip 4.0 cm
\centerline {{\bf {FIGURE CAPTIONS}}}
\vskip 0.5 cm
\item{{\bf Fig.1}} The solid line I=0,l=0 pion scattering phase shift
calculated from
unitarized CPTh. Dashed line corresponds to the similar phase shift where the
left hand cut is neglected. The dot dashed line is the CPTh prediction.
 The experimental data are those of Estabrook et al. Ref.[7].\hfill\break

\item{{\bf Fig.2}} The solid line I=1,l=1 pion scattering phase shift
calculated from
the unitarized CPTh. Dashed line corresponds to the similar phase shift
where the
left hand cut is neglected. The dot dashed line is the CPTh prediction.
 The experimental data are those of Estabrook et al. Ref.[7]. The
difference between the
solid and the dashed line is not distinguishable.\hfill\break

\item{{\bf Fig.3}}The solid line is the pion vector form factor squared
calculated from the
Omn\'es representation. The dashed line represents the same quantity
calculated by the approximate solution
where the left cut contribution is neglected. The experimental data are those
of
L.M. Barkov et al Ref[14]. The difference between the
solid and the dashed line is not distinguishable.\hfill\break

\item{{\bf Fig.4}} The solid line is the pion scalar form factor squared
calculated from the
Omn\'es representation. The dashed line represents the same quantity
calculated by the approximate solution
where the left cut contribution is neglected.\hfill\break

\end